\documentstyle[12pt,damtp]{article}
\title{Signature Characters for ${\rm A}_2$ and ${\rm B}_2$}
\author{Adrian Kent
 and G\'{e}rard Watts\damtpduritp}
\abstract{
The signatures of the inner product matrices on a Lie algebra's
highest weight representation are encoded in the representation's
signature character.  We show that the signature characters of a
finite-dimensional Lie algebra's highest weight representations obey
simple difference equations that have a unique solution once
appropriate boundary conditions are imposed.  We use these results to
derive the signature characters of all $A_2$ and $B_2$ highest weight
representations.
Our results extend, and explain, signature
patterns analogous to those observed by Friedan, Qiu and Shenker in
the Virasoro algebra's representation theory.}

\begin{document}
\maketitle

\input mssymb
\newenvironment{proof}{{\bf Proof}}{}
\newtheorem{lemma}{Lemma}
\newtheorem{theorem}[lemma]{Theorem}
\newtheorem{corollary}[lemma]{Corollary}
\def\floor#1{\lfloor#1\rfloor}
\def\blank#1{}
\def\swap#1#2{#1}
\def \ul {\underline}
\def \eps {\epsilon}
\def\ut#1{\hbox{\boldmath #1}}
\def\nn{\nonumber\\}
\def\tr{{\rm tr}}
\def\fif{{\rm~if~}}
\def\num{{\rm num}}
\def\sgn{{\rm sgn}}
\def\sig{{\rm sig}}
\def\diag{{\rm diag}}
\def\for{{\rm ~for~}}
\def\pol{{\rm pol}}
\def\quart{\frac{1}{4}}
\def\half#1{\frac{#1}{2}}
\def \hb {h_\beta}
\def \ha {h_\alpha}
\def \hbd {$\hb$}
\def \had {$\ha$}

\def\Ep{E_+}
\def \Ea  {E_{\alpha}}
\def \Eb  {E_{\beta}}
\def \Eab {E_{\alpha+\beta}}
\def\Em{E_-}
\def \Ema  {E_{-\alpha}}
\def \Emb  {E_{-\beta}}
\def \Emab {E_{-\alpha-\beta}}

\section{Introduction}

The theory of non-unitary highest weight Lie algebra representations
is still relatively undeveloped.  An interesting new approach was
suggested by the work of Friedan, Qiu and Shenker (FQS) who, in one of
the foundational papers \cite{fqs} of conformal field theory, used the
Virasoro algebra's determinant formula \cite{kac,ff} to analyse the
unitarity of its highest weight representations.  They established the
unitarity of a continuum of the representations, and the non-unitarity
of all other representations except for an infinite discrete series.
Drawing on evidence from statistical physics, as well as computation,
they conjectured that the representations in the discrete series are
unitary. This was later proven by another method \cite{gko}.

The relevance of FQS's work to the representation theory of
finite-\linebreak dimensional Lie algebras is not immediately obvious,
since the unitary highest weight representations of these algebras can
be classified simply by studying the induced representations of the
embedded $su(2)$ subalgebras.  However, as FQS recognised, the
emergence of the discrete series suggests very strongly that there are
definite patterns in the dependence of the inner product matrix
signatures on a representation's location amongst the vanishing
curves, and that the unitarity of the discrete series can be
understood as a consequence of these patterns.  Thus one should look
for results that describe and explain these patterns and hence give
explicit expressions for all inner product matrix signatures of all
highest weight representations.

Here, we carry this through for the Lie algebras $A_2$ and $B_2$.
(These are the simplest non-trivial cases, since the relevant vector
spaces for $A_1$ are one-dimensional and their inner product matrix
signatures are easily calculable.)  We obtain explicit (and
intriguingly simple) expressions for these algebras' signature
characters (generating functions in which the inner product matrix
signatures are encoded).  Our arguments apply to many other cases of
interest, including all finite-dimensional Lie algebras.

\section{Notation and Basic Results}

We follow the conventions of Kac and Kazhdan \cite{kk} with
modifications.  Let $g$ be a complex semisimple Lie algebra of rank
$r$ with symmetrisable Cartan matrix $A=(a_{ij})$, that is,
\mbox{$A = D A_{\rm sym}$}, where $D=\diag( d_1 , \ldots , d_r )$,
\mbox{$A_{\rm sym} = (a_{ij}^{\rm sym})$}
is symmetric and the $d_i$ are non-zero.  Let $h$ be a Cartan
subalgebra of $g$; let $\{ \alpha_1 , \ldots , \alpha_r
\}$ be a basis of simple roots with respect to $h$; $\Delta_+$ the set
of positive roots with respect to this basis; $\Delta_-$ the set of
negative roots.  Define the positive and negative root semilattices to
be $\Lambda_{\pm} = \{ \pm \sum_{i=1}^r n_i \alpha_i : n_i \in \Bbb
Z_{+} \}$.  Let $g_\alpha$ be the root subspace of $g$ corresponding
to the root $\alpha$, and define the subalgebras $n_+ = \sum_{\alpha
\in \Delta_+} g_\alpha$, $n_- = \sum_{\alpha \in \Delta_-} g_\alpha$,
so that $g = n_- \oplus h \oplus n_+$.  We choose a set of generators
$\{ E_{\alpha_i} , E_{- \alpha_i}, H_i : 1 \leq i \leq r \}$ for $g$;
here $\{ H_1 , \ldots , H_r \}$ is a basis for $h$, and $E_{\pm
\alpha_i}
\in g_{\pm \alpha_i}$.  The Killing form $\iptwo{ \,}{ \, }$ on $g$ is the
unique symmetric invariant non-degenerate bilinear form on $g$ with
the property that $\iptwo{ H_i }{ H_j } = a_{ij}^{\rm sym}$.  The
Killing form, restricted to $h$, induces a form on $h^*$ which we also
denote by $\iptwo{ \, }{ \, }$.  We have $\iptwo{ \alpha_i }{ \alpha_j
} = a_{ij}^{\rm sym}$ and $a_{ij} = 2 \iptwo{ \alpha_i }{ \alpha_j } /
\iptwo{ \alpha_i }{ \alpha_i }$.  Finally, we define $\rho \in h^*$ by
$\iptwo{ \rho }{\alpha_i} = \frac{1}{2} \iptwo{\alpha_i}{\alpha_i} $
for $\alpha_i$ a simple root.  The generators obey the relations
\eq
\begin{array}{rcl}
{}~[ H_i , H_j \hfil ] \, \, & = & 0 \, , \\ ~[ H_i , E_{\pm \alpha_j }
] \, & = & \pm a_{ij} E_{\pm \alpha_j } \, , \\ ~[ E_{ \alpha_i} ,
E_{-\alpha_j} ] & = &
\delta_{ij}
\frac{2 \iptwo{ E_{\alpha_i}}{ E_{- \alpha_i}}}{\iptwo{\alpha_i}{\alpha_i}}
\,H_i \, .
\end{array}
\en

Denote by $U(a)$ the universal enveloping algebra of the Lie algebra
$a$.  For $\lambda \in h^* $, the Verma module representation $V(
\lambda )$ of $g$ is the representation of $g$ containing a vector
$\ket{ \lambda }$ such that
\eq
U(n_+ ) \ket{ \lambda } = 0 \,,\; H_i \ket{ \lambda } =
\frac{ 2 \iptwo{ \alpha_i \, }{ \, \lambda }}
{ \iptwo{ \alpha_i \, }{ \, \alpha_i }}
\ket{ \lambda } \, ,
\en
and such that if $x,y \in U( n_- )$ then $x \ket{ \lambda } = y \ket{
\lambda }$ only if $x=y$.
We have the decomposition $V( \lambda ) = \oplus_{\mu} ( V ( \lambda )
)_{\mu}$, where
\eq
V( \lambda )_{\mu} = \{ v \in V( \lambda ) : H_i v = \frac{2
\iptwo{\alpha_i }{ \lambda - \mu}}{\iptwo{\alpha_i}{\alpha_i}}
v
\} \, .
\en
We consider below representations $V(
\lambda )$ for $\lambda$ such that $\iptwo{ \alpha_i \, }{ \, \lambda }$ is
real.

Define an adjoint map $\dagger$ on $g$ to be the unique algebra
anti-automorphism $\dagger : g \rightarrow g$ such that $(H_i
)^{\dagger} = H_i $, $(E_{\alpha_i} )^{\dagger} = E_{- \alpha_i}$ and
$(E_{- \alpha_i} )^{\dagger} = E_{\alpha_i}$, and extend this to an
algebra anti-automorphism on $U( g )$, also denoted by $\dagger$.  The
Shapovalov form on $V(\lambda)$ is the unique bilinear form
$\ipone{\,}{\,}$ such that $\ipone{\lambda}{\lambda} = 1$ and
$\ipone{x
\lambda}{ y \lambda } = \ipone{ \lambda}{ (x )^{\dagger} y \lambda} =
\ipone{( y )^{\dagger} x \lambda}{\lambda}$ for $x, y \in U(g)$.
We have that $ \ipone{V(\lambda)_{\mu}}{ V(\lambda)_{\nu}} = 0$ if $
\mu \neq \nu$.  Denote by $M_{\mu} (\lambda )$ the real symmetric
matrix defining (in some choice of basis) the inner product restricted
to $V(\lambda)_{\mu}$.  For $\lambda \in h^*$ we write $\lambda = -
\sum_{i=1}^{r} \lambda^i
\alpha_i$.

Now the character of $V( \lambda )$ can be defined as
\eq
\chi (V( \lambda )) =  ( x_1 )^{\lambda^1} \ldots ( x_r )^{\lambda^r}
                        \sum_{\mu \in \Lambda_-} \dim (V(\lambda
)_{\mu} ) ( x_1 )^{\mu^1} \ldots ( x_r )^{\mu^r} \, ,
\en
In physicists' notation, defining the fundamental co-weights $h_i \in
h$ so that $\alpha_j (h_i ) =
\delta_{ij}$, letting $x_i = \exp ( - i \theta_i )$, $\theta =
(\theta_1 , \ldots , \theta_r )$ and $H = ( h_1 ,
\ldots , h_r )$, and writing ${\theta}\cdd{H} =
\sum_{i=1}^r \theta_i h_i$, we have
\eq
\chi (V( \lambda )) =\Tr_{V( \lambda )} ( e^{i {\theta}\cdd{H}}) \, .
\en

By analogy, the {\it signature character} of $V( \lambda )$ is defined
as
\eq
\chi^{\rm sig} (V( \lambda )) =
( x_1 )^{\lambda^1} \ldots ( x_r )^{\lambda^r}
\sum_{\mu \in \Lambda_-} {\rm sig} (M_{\mu} (\lambda ) )
                       ( x_1 )^{\mu^1} \ldots ( x_r )^{\mu^r} \, ,
\en
where ${\rm sig}(M)$ denotes the signature of the matrix $M$; that is,
if $M = S D S^T$, where $S$ is non-singular and $D = {\rm diag}(+1,
\ldots, +1, 0, \ldots , 0, -1 , \ldots , -1 )$, then ${\rm sig}(M) =
{\rm tr} (D)$.  This can be rewritten by defining a linear operator
$P$ on $V( \lambda )$ with the property that if $v \in (V( \lambda
))_{\mu}$ is an eigenvector of $(M_{\mu} (\lambda ) ) $ with
eigenvalue $a_v$ then
\eq
P v = \sgn ( a_v ) v \, ,
\en
where the sign function is given by
\eq
\sgn ( t ) = \left\{
\begin{array}{ll}
1 & \for t > 0 \, , \\ 0 & \for t = 0 \, , \\ -1 & \for t < 0 \, .
\end{array}
\right.
\en
Then
\eq
\chi^{\rm sig} (V( \lambda )) =
\Tr_{V( \lambda )} (P  e^{i {\theta}\cdd{H}}) \, .
\en

We shall work with normalised signature characters
\eq
\sigma ( \lambda ) \equiv
e^{-i\theta \cdd H (\lambda) } \chi^{\rm sig} (V( \lambda ))
\, .
\en
The fundamental result we shall need is Shapovalov's determinant
formula \cite{shap,kk},
\eq
\label{shap}
\det (M_{\mu} (\lambda )) = C \prod_{\alpha \in \Delta_+} \prod_{n>0}
((\alpha,\lambda+\rho)- n \frac{\iptwo{\alpha}{\alpha}}{2})^{P( \mu +
n \alpha )} \, ,
\en
where if the weight decomposition of $U( n_- )$ is given by
$\oplus_{\mu } U( n_- )_{\mu}$ then $P(\mu) = \dim ( U( n_- )_{\mu}
)$, and $C$ is a non-zero basis-dependent constant.  We call the
planes $P(n, \alpha ) \equiv \{ \lambda \in h^* :
(\alpha,\lambda+\rho)=n \}$ the
\mbox{\it Shapovalov vanishing planes} of $g$.

\section{General properties of
signature characters for finite-dimensional Lie algebras}

For the remainder of the paper we take $g$ to be a finite-dimensional
Lie algebra.  Since the matrices $M_{\mu}(\lambda )$ are real and
symmetric, we have that
\eq
{\rm sig} ( M_{\mu} (\lambda ) ) =
\sum_{j=1}^{P( \mu )} {\rm sgn}(t_j ) \, ,
\en
where $\{ t_1 , \ldots , t_{P( \mu )} \}$ are the eigenvalues of
$M_{\mu} (\lambda )$.  It follows that ${ \rm sig} ( M_{\mu} (\lambda
) ) $ is constant on any connected region of $h^*$ in which $\det (
M_{\mu} (\lambda ) ) \neq 0$, and in particular that $\sigma (V(
\lambda ))$ is constant on connected regions $R$ in which $\det (
M_{\mu} (\lambda ) ) \neq 0$ for any $\mu \in \Lambda_-$ and any
$\lambda \in R$.  Our first result is the observation that the change
of the signature characters between two such neighbouring regions is
given by a simple difference equation.

We consider a point $\lambda_0 \in h^*$ lying on precisely one of the
vanishing planes $P( n_0 , \alpha_0 )$.  Choose a basis $\{ \kappa_i :
1 \leq i \leq r \}$ for $h^*$ such that $( \kappa_i , \alpha_0 ) = 0 $
for $ 1 \leq i \leq (r-1)$ and $ \kappa_r = \alpha_0 $, and define
coordinates
\mbox{$( t_1 , \ldots , t_{r-1}, \epsilon ) \equiv ( \ul{t} , \epsilon )$}
for $\lambda \in h^*$ by the equation $\lambda = \lambda_0 +
\sum_{i=1}^{r-1} t_i \kappa_i +
\epsilon \kappa_r$.
We write $V ( \ul{t} , \epsilon )$ for $V( \lambda )$ and $\ket{
\ul{t} , \epsilon }$ for $\ket{\lambda }$, and we adopt the convention
that any vector written in the form $w ( \ul{t} , \eps )$ belongs to
$V ( \ul{t} , \epsilon )$.  Define the neighbourhood $N ( \delta )$ of
$\lambda_0$ as the coordinate region with $ | \ul{t} | < \delta$ and $
| \epsilon | < \delta$, and choose $\delta_0$ sufficiently small such
that $N ( \delta_0 )$ intersects no vanishing plane other than $P( n_0
, \alpha_0 )$.  Shapovalov's formula implies that, for $| \ul{t} | <
\delta_0$, the Verma module $V ( \ul{t} , 0 )$ contains a unique
descendent highest weight vector $v( \ul{t} , 0 ) \in V ( \ul{t} , 0
)_{n_0 \alpha_0}$.  We can express $v( \ul{t} , 0 )$ as $a( \ul{t} )
\ket{ \ul{t} , 0 }$, where $a( \ul{t} ) \in U( n_- )$ is defined for $
| \ul{t} | < \delta_ 0$.  Define the vectors $v( \ul{t} , \eps ) \in V
( \ul{t} , \eps )_{n_0 \alpha_0}$ by $v( \ul{t} , \eps ) = a( \ul{t} )
\ket{ \ul{t} , \eps}$,

If $\mu + n_0 \alpha_0 \notin \Lambda_- $ then the inner product
matrix $M_{\mu} (\lambda )$ is non-singular in $N ( \delta_0 )$ and
its signature is constant throughout the neighbourhood.  Now we
consider $M_{\mu} (\lambda )$ for $\mu$ such that $\mu + n_0 \alpha_0
\in \Lambda_- $.  Reducing $\delta_0$ if necessary, we choose $a_i \in
(U( n_- ))_{\mu + n_0 \alpha_0} $ for $ 1 \leq i \leq P( \mu + n_0
\alpha_0 )$ and $b_i \in (U( n_- ))_{\mu}$ for $P( \mu + n_0 \alpha_0
) + 1 \leq i \leq P( \mu )$ such that, setting
\eq
v_i ( \ul{t} , \epsilon ) = \left\{
\begin{array}{ll}
a_i v( \ul{t} , \eps ) & \for 1 \leq i \leq P( \mu + n_0 \alpha_0 ) \,
, \\ b_i \ket{\ul{t} , \eps } & \for P( \mu + n_0 \alpha_0 ) + 1 \leq
i
\leq P( \mu )  \, ,
\end{array}
\right.
\en
the set $\{ v_i ( \ul{t} , \epsilon ) : 1 \leq i \leq r \}$ forms a
basis of $V ( \ul{t} , \eps )_{\mu}$ throughout $N( \delta_0 )$.

Now
\eq
\ipone{v_i ( \ul{t},\eps )}{v_j ( \ul{t} ,\eps )} = O(\eps )
\en
for $1\le i\le P( \mu + n_0 \alpha_0 ) $ or $1\le j \le P( \mu + n_0
\alpha_0 )$.  Thus in this basis the inner product matrix $M_{\mu}$
has the form
\eq
 \left( \begin{array}{cc} M' & X \\ X^T & M'' \end{array} \right) \;,
\en
where
\eq
\begin{array}{rcll}
(M')_{ij} & = & M_{ij} & \for 1\le i,j\le P( \mu + n_0 \alpha_0 ) \, ,
\\ (M'')_{ij} & = & M_{i+P( \mu + n_0 \alpha_0 ),j+P( \mu + n_0
\alpha_0 )} &
\for 1 \le i,j \le P( \mu ) - P( \mu + n_0 \alpha_0  ) \, ,
\end{array}
\en
and $M^{\prime} ( \ul{t},\eps )$ and $X ( \ul{t},\eps )$ are $O(\eps
)$ in the neighbourhood $N ( \delta_0 )$.

For $( \ul{t},\eps ) \in N ( \delta_0 )$, equation (\ref{shap})
implies that
\eq
\det ( M ( \ul{t},\eps ) ) = A( \ul{t} ) \eps^{P(\mu + n_0 \alpha_0  )} +
                                O ( \eps^{P( \mu + n_0 \alpha_0 )+1} )
\en
for some non-zero function $A( \ul{t} )$.  As $M'' ( \ul{t},\eps ) =
M'' ( \ul{t},0 ) + O( \eps )$ and \mbox{$\det ( M ( \ul{t},\eps ) )$}
is non-zero for $( \ul{t},\eps ) \in N ( \delta_0 )$ if $ \eps \neq
0$, $M'' ( \ul{t},0 )$ must be non-singular.  Hence, again reducing
$\delta_0 $ if necessary, we can assume $M''( \ul{t} , \eps )$ is
non-singular throughout $N( \delta_0 )$.

Thus there is a new basis in which $M_{\mu}$ has the form
\eq
 \left( \begin{array}{cc} M' - X (M'' )^{-1} X^T & 0 \\ 0 & M''
\end{array} \right) \, .
\en
Likewise \mbox{$M'''( \ul{t} ) \equiv \lim_{\eps \rightarrow 0}
\frac{1}{\eps} M^{'} ( \ul{t} , \eps )$} must be non-singular.
Hence, for sufficiently small $\eps$,
\eq
\label{cross}
\begin{array}{rcl}
\sig(M( \ul{t} , \eps )) &= &
\sig(M'( \ul{t} , \eps )) + \sig(M''( \ul{t} , \eps )) \\
&= & \sgn( \eps ) \sig(M'''(\ul{t} )) + \sig(M''(\ul{t},0)) \, .
\end{array}
\en

Now, the Poincar\'e-Birkhoff-Witt theorem implies that $( a_i
)^{\dagger} a_j$ can be expressed as a sum of products $\sum_k r_k s_k
t_k$ with $r_k \in U( n_- )$, $s_k \in U(h)$ and $t_k
\in U( n_+ )$.
Given such an expression, we define $H_{ij} = \sum_{\{ k: r_k = t_k =
1\} } s_k $.  The definition is in fact independent of the expression
chosen, since $U(n_{\pm})$ and $U( h)$ are mutually disjoint
subalgebras.  Now if $a \in n_+$ then $ a \, v(\ul{t},0)=0$ and so $ a
\, v(\ul{t},\eps ) = O(\eps )$.  Hence
\eq
\ipone{v_i (\ul{t},\eps )}{v_j (\ul{t},\eps )} =
\ipone{v(\ul{t},\eps )}{H_{ij} v(\ul{t},\eps )} + O(\eps^2) \, .
\en

Thus
\eq
\eps M^{'''}( \ul{t} ) =
\ipone{v( \ul{t} , \eps )}{v( \ul{t} , \eps )}
                       M_{\mu + n_0 \alpha_0} ( \lambda_0 - n_0
\alpha_0 ) + O ( \eps^2 ) \, ,
\en
where the latter inner product matrix is taken in the basis \newline
$\{ a_i \ket{ \lambda_0 - n_0 \alpha_0 } :
1 \leq i \leq P ( \mu + n_0 \alpha_0 ) \}$.

This establishes the following result:

\begin{theorem} Let $g$, $h$, $\lambda_0$, be as above.
Then, in the notation previously defined
\eq
\label{scde}
\lim_{\eps \rightarrow 0^+} \sigma(\ul{t},\eps )  -
\lim_{\eps \rightarrow 0^-} \sigma(\ul{t},\eps )  =
	2 \, \sgn(
\ipone{v( \ul{t} , \eps )}{v( \ul{t} , \eps )} |_{\eps=0^+} )
 e^{-in_0 \theta \cdd H ( \alpha_0 )}
\sigma( \lambda - n_0 \alpha_0 ) \, .
\en
\end{theorem}

This, the main result of the section, we refer to as the {\it
signature character difference equation}.  We also have, from equation
(\ref{cross}):

\begin{lemma}
\label{l.average}
Let $g$, $h$, $\lambda_0$, be as above.  Then, in the notation
previously defined,
\eq
\label{average}
\sigma(\lambda_0 ) = \frac{1}{2}
( \lim_{\eps \rightarrow 0^+} \sigma(\ul{t},\eps ) +
\lim_{\eps \rightarrow 0^-} \sigma(\ul{t},\eps ) ) \, .
\en
\end{lemma}

The next result requires some preliminary notation.  Suppose that a
formal power series valued function $f:R\rightarrow \Bbb C [[
e^{-i\theta_1},\ldots	e^{-i\theta_r} ]] $ is defined on a subset $R$
of $h^*$ that includes almost all the points $t \rho$ and $-t \rho$
for real $t$.  Define functions $ f_{ \alpha }$ on $R$ by $f ( \lambda
) = \sum_{ \alpha \in \Lambda_- } f_{ \alpha } (\lambda )
e^{i\theta \cdd H ( \alpha )} $ for $\lambda \in R$.  Then we define
$\lim_{\lambda \rightarrow \infty} f( \lambda )$ to be
\eq
 \sum_{\alpha \in \Lambda_- } \lim_{t \rightarrow \infty} f_{ \alpha }
( t \rho )
e^{i\theta \cdd H ( \alpha ) } \, ,
\en
whenever these latter limits exist.  We define $\lim_{\lambda
\rightarrow - \infty} f( \lambda )$ similarly.  A partial ordering is
defined on $\Lambda_-$ by setting $\alpha < \beta$ if $\beta - \alpha
\in \Lambda_+$.

\begin{theorem}
\label{guessright}
Let $g$ be a finite-dimensional Lie algebra of rank $r$; let $h$ be a
Cartan subalgebra of $g$; let $R$ be the subset of
$h^*$ comprising all points $\lambda$ for which no Shapovalov
determinant $\det (M_{\mu} (\lambda ))$ vanishes.  Let $f:R
\rightarrow \Bbb C [[ e^{-i\theta_1},\ldots,e^{-i\theta_r} ]]$ be a
power series valued function on $R$ with the properties that \newline
(i) $\lim_{\lambda \rightarrow - \infty} f( \lambda ) = \lim_{\lambda
\rightarrow - \infty} \sigma ( \lambda )$;\newline (ii) $\lim_{\lambda
\rightarrow \infty} f( \lambda ) = \lim_{\lambda \rightarrow \infty}
\sigma ( \lambda )$;\newline (iii) if $\lambda_0 $ is a point on
precisely one of the Shapovalov vanishing planes $P( n_0 , \alpha_0
)$, and $(\ul{t},\eps )$ are the coordinates defined above in a
neighbourhood $N ( \delta )$ of $\lambda$ which intersects precisely
one vanishing plane, then $f$ obeys the equation
\eq
\label{pmscde}
\lim_{\eps \rightarrow 0^+} f(\ul{t},\eps )  -
\lim_{\eps \rightarrow 0^-} f(\ul{t},\eps )  =
        \pm 2 e^{-in_0 \theta \cdd H (\alpha_0 )}
            f( \lambda_0 - n_0 \alpha_0 )  \, ;
\en
(iv) Each function $f_{ \alpha }$ is constant on the union of any
connected components of $R$ that are not separated by any vanishing
plane $P( n , \beta)$ with $ n \beta < \alpha$.

Then $f( \lambda ) = \sigma ( \lambda )$ for all $\lambda \in R$.
\end{theorem}

\begin{proof}
For suppose $f \neq \sigma$.  Let $ \alpha $ be such that $f_{ \alpha
} (\lambda ) \neq \sigma_{ \alpha } (\lambda )$ for some $\lambda \in
R$ and such that if $f_{ \beta } (\lambda ' ) \neq
\sigma_{ \beta } (\lambda ' )$ for
some $\lambda ' \in N$ then $\beta \geq \alpha$.
We have the component form of equation (\ref{pmscde}):
\eq
\label{comp}
\lim_{\eps \rightarrow 0^+} f_{\alpha}(\ul{t},\eps )  -
\lim_{\eps \rightarrow 0^-} f_{\alpha}(\ul{t},\eps )  =
        \pm 2 f_{\alpha + n_0 \alpha_0} ( \lambda_0 - n_0 \alpha_0 )
\, .
\en
Now, since $\lim_{\lambda \rightarrow \infty} f_\alpha(\lambda) =
\lim_{\lambda \rightarrow \infty} \sigma_\alpha(\lambda)$,
there must be some point $\lambda_0$ that lies on precisely one
Shapovalov vanishing plane $P(n_0 ,
\alpha_0 )$ and such that in the neighbourhood of $\lambda_0$ the
functions $f_{\alpha}$ and $\sigma_\alpha$ satisfy
equation (\ref{comp})
with opposite signs on the right hand side.
Since the same sign
holds for all components of $f$ crossing the plane $P(n_0,\alpha_0)$ at
$\lambda_0$,
and since properties (i) and (iii) imply that $f_0 (\lambda ) = 1$ for all
$\lambda$,
we have that
\eq
\lim_{\eps \rightarrow 0^+} f_{n_0\alpha_0}(\ul{t},\eps )  -
\lim_{\eps \rightarrow 0^-} f_{n_0\alpha_0}(\ul{t},\eps )  =
        \pm 2
\, ,
\en
where the sign on the right hand side is the opposite to that in the
analogous equation satisfied by $\sigma_{n_0\alpha_0}$.
So since $f_{\beta} = \sigma_{\beta}$ for all $\beta < \alpha$, we must have
that $n_0 \alpha_0 = \alpha$, and that $f$ satisfies equation (\ref{scde})
across all sections of all vanishing planes $P( n, \beta )$ such that
$n \beta < \alpha$. Now $f_{ \alpha } (\lambda ) =
\sigma_{ \alpha } (\lambda )$ asympotically as
$\lambda \rightarrow \pm \infty$.  Moreover, by definition,
$f_{\alpha}$ and $\sigma_{\alpha}$ are constant on any set of regions
in $R$ that are not separated by a plane $P( n , \beta )$ with $n
\beta < \alpha$.  But a well known property of the root systems of
finite-dimensional Lie algebras is that if $\alpha$ and $ \beta$ are
positive roots and $m$ and $n$ are positive integers such that $m
\alpha = n \beta$ then $m = n$ and $\alpha = \beta$.  Hence $P(n_0 ,
\alpha_0 ) $ is the only vanishing plane $P(n, \beta )$ such that $n
\beta = n_0 \alpha_0$.  But now we have established that $f_{ \alpha }
= \sigma_{ \alpha }$ for all points in $R$ above $P(n_0 , \alpha_0 )
$, and likewise for all points in $R$ below $P(n_0 , \alpha_0 ) $.
Thus $f_{ \alpha } = \sigma_{ \alpha }$ throughout $R$, which is a
contradiction.
\end{proof}
\newpage
We now obtain explicitly the asymptotic signature characters for
all semi-simple finite dimensional Lie algebras $g$.

\begin{lemma}
\label{l.asymp}
Let $\sigma(\lambda)$ be the signature character of the
highest weight representation $V( \lambda )$ of the Lie algebra $g$.
Then, with the above definitions,
\eq
\label{l.achar}
\begin{array}{rcl}
{\displaystyle \lim_{ \lambda \rightarrow \infty}} \sigma (\lambda ) &=&
{\displaystyle\prod_{\alpha\in\Delta_+}}
(1 - e^{-i \theta \cdd H ( \alpha )})^{-1} \\
{\displaystyle \lim_{\lambda \rightarrow - \infty}} \sigma (\lambda ) &=&
{\displaystyle\prod_{\alpha\in\Delta_+}}
(1 + e^{-i \theta \cdd H ( \alpha )})^{-1} \, .
\end{array}
\en
\end{lemma}

\begin{proof}
By the Poincar\'e-Birkhoff-Witt theorem, we may take a basis of
$U(n_- )_\mu$ to be lexicographically ordered monomials in the lowering
operators of $g$. Each monomial corresponds to some ordered partition
\mbox{$\{\beta\} = \{ \beta_1 , \ldots , \beta_k \}$}
of $\mu$ in terms of negative
roots. We denote the set of such partitions by
$\Pi(\mu)$.
Take generators $E_{\alpha} \in g_{\alpha}$.
We consider the inner product matrix
$M_{\mu} (\lambda )$ in the basis
$ \{ v_{i} (\lambda ) : 1 \leq i \leq P(\mu) \} $, where the vector
$v_i$ corresponds to the $i$-th partition $\{\beta_1,\ldots,\beta_k\}$
of $\mu$, and
\eq
v_i  =  E_{\beta_1}\ldots E_{\beta_k} \ket{\lambda} \,.
\en
As $\lambda \rightarrow \pm \infty$, $M_{\mu} (\lambda )$
is asymptotically diagonalised, in the sense that
\eq
\lim_{\lambda \rightarrow \pm \infty }
\sig ( M_{\mu} (\lambda ) ) =
\sum_{i=1}^{P(\mu)} \lim_{\lambda \rightarrow \pm \infty }
\sgn ( M_{\mu} (\lambda )_{ii} ) \, .
\en
Now
\eq
\lim_{t \rightarrow \infty}
\sgn ( \ipone{ v_{\{\beta\}} (\pm t \rho)}{v_{\{\beta \}} (\pm t \rho )} ) =
 (\pm 1)^{l(\{\beta\})} \, ,
\en
where the length $l( \{\beta\} )$ of a partition is the number of roots
which appear in the partition.
Hence
\eq
\lim_{\lambda \rightarrow \pm\infty }
\sig ( M_{\mu} (\lambda ) ) = \sum_{\{\beta\} \in \Pi(\mu)}
(\pm 1)^{l(\{\beta\})} \,.
\en
Thus the signature characters $\sigma(\lambda)$ obey
\eq
\begin{array}{rcl}
{\displaystyle \lim_{\lambda \rightarrow \pm \infty}}\sigma(\lambda) &=&
{\displaystyle\sum_{\mu\in\Lambda_-}} \, \,
{\displaystyle\sum_{\{\beta\}\in\Pi(\mu)}} (\pm 1)^{l(\{\beta\})}
e^{i\theta \cdd H ( \mu ) }  \\
	&=& {\displaystyle\prod_{\alpha\in\Delta_+}}
	(1 \mp e^{-i\theta \cdd H ( \alpha )})^{-1} \, .
\end{array}
\en
This completes the proof.
\end{proof}

\section{Signature characters of $A_2$}

We now use the results of the last section to obtain signature characters
for the highest weight representations of $A_2$.  In the notation defined
above, the $A_2$ Cartan matrix is $A = \left( \begin{array}{rr} 2 & -1 \\ -1 &
2 \end{array} \right)$ and the positive root set $\Delta_+ = \{
\alpha_1 , \alpha_2 ,
\alpha_3=\alpha_1 + \alpha_2 \}$.
The $A_2$ partition function is
\eq
P ( \mu ) = \left\{
\begin{array}{ll}
\min ( m + 1 , n + 1 )  &  {\rm if~} \mu = - m \alpha_1 - n \alpha_2 \for
{\rm integers~}m {\rm~and~} n \geq 0 \, ,  \\ 0 & {\rm otherwise.}
\end{array}
\right.
\en

{}From the Shapovalov formula (\ref{shap}), it follows that $R$
splits into a union of connected subregions:
\eq
R \, \, = \bigcup_{(a_1 , a_2 , a_3 ) \in I} R(a_1 , a_2 , a_3 ) \, ,
\en
where
\eq
\begin{array}{rl}
I = \{ (a_1 , a_2 , a_3 ) : a_i \hbox{ ~non-negative~integers~} &
{\rm~with~} a_1 + a_2 = a_3 \\
& {\rm ~or~}a_1 + a_2 + 1 = a_3 \\
& {\rm ~or~} a_1 = 0{\rm~and~}a_2 > a_3 \\
& {\rm ~or~}a_2 = 0{\rm ~and~}a_1 > a_3 \}
\end{array}
\en
and
\eq
R(a_1 , a_2 , a_3 ) = \{ \lambda \in h^* :
	a_i = \max(0,\floor{(\alpha_i,\lambda+\rho)}) \} \, .
\en

We set $e^{-i \theta_j} = x_j$ for $j=1,2$.
We shall define a formal power series
valued function $f: R \rightarrow \Bbb C [[ x_1 , x_2 ]]$ that is
constant on the connected components $R(a_1 , a_2 , a_3 )$ of $R$ and
is asympotically equal to the signature character
$\sigma ( \lambda)$ of an $A_2$ highest weight representation $V (
\lambda ) $ as \mbox{$\lambda \rightarrow \pm \infty$}.  We then
show that $f$ satisfies equation (\ref{pmscde}).  It will then follow,
from theorem \ref{guessright}, that $f(\lambda )$ is precisely the signature
character $\sigma ( \lambda )$ for $\lambda \in R$.

We define $f$ on $R$ in terms of a function $\tilde{f}$ defined on
$I$, so that
\eq
\label{fdef}
f ( \lambda ) =
\frac{\tilde{f} ( a_1 , a_2 , a_3 )}{(1-x_1^2 )(1 - x_2^2 )
( 1 - (x_1 x_2 )^2 )} {\rm~if~} \lambda \in
R(a_1 , a_2 , a_3 ) \, .
\en
We take
\eq
\label{gdef}
\begin{array}{rcl}
\tilde{f} ( a_1 , a_2 , a_3 ) & = &
{}~~(1 + x_1 )(1 + x_2 )(1+ x_1 x_2 ) \\
&&-    2 (x_1 )^{a_1 + 1} (1 + x_2 )(1+ x_1 x_2 ) \\
&&-                    2 (x_2 )^{a_2 + 1} (1 + x_1 )(1+ x_1 x_2 )\\
&&-                 2 (x_1 x_2)^{a_3 + 1} (1 + x_1 )(1+ x_2 ) \\
&&+ 4 (x_1)^{a_1 + 1} (x_1 x_2 )^{\min (a_2 , a_3 ) + 1 } (1 + x_2 ) \\
&&+ 4 (x_2)^{a_2 + 1} (x_1 x_2 )^{\min (a_1 , a_3 ) + 1 } (1 + x_1 ) \\
&&+ 4 (x_1)^{\max(a_1 ,a_3 ) + 1} (x_2 )^{\max (a_2 , a_3 ) + 1 }
     (1 + x_1 x_2 )\\
&&-  8 (x_1 )^{\min(a_2 , a_3 ) + a_1 + 2} (x_2 )^{\min(a_1 , a_3 ) + a_2 + 2}
   \, .
\end{array}
\en
The following two results are easily verified from equations
(\ref{fdef}-\ref{gdef}).
\begin{lemma}
\label{l.a2pmscde}
With the above definitions and conventions, let $\lambda_0$ be a point lying on
precisely one $A_2$ vanishing line $P( n_0 , \alpha_0)$.  Then
\eq
\lim_{\eps \rightarrow 0^+} f(\ul{t},\eps )  -
\lim_{\eps \rightarrow 0^-} f(\ul{t},\eps )  =
        e(\lambda_0 ) 2 x_1^{ - n_0 \alpha_0^1 } x_2^{ - n_0 \alpha_0^2 }
        f( \lambda_0 - n_0 \alpha_0 ) \, ;
\en
where the sign function $e(\lambda_0 )$ is given by
\eq
e(\lambda_0 ) = \left\{
\begin{array}{ll}
+1 & \fif \alpha_0 = \alpha_1 {\rm ~or~} \alpha_2 \, , \\
+1 & \fif \alpha_0 = \alpha_3 {\rm ~and~} \lambda_0 + \rho \notin C \, , \\
-1 & \fif \alpha_0 = \alpha_3 {\rm~and~} \lambda + \rho \in C \, ,
\end{array}
\right.
\en
and where $C$ is the fundamental Weyl chamber of $A_2$, that is,\newline
$C = \{ \lambda \in h^* : \iptwo{ \alpha_1 }{ h } \geq 0 {\rm~and~}
                          \iptwo{ \alpha_2 }{ h } \geq 0 \}$.
\end{lemma}

\begin{lemma}
\label{l.a2asy}
With the above definitions,
\eq
\begin{array}{rcl}
lim_{\lambda \rightarrow \infty} f (\lambda ) &=&
( 1 - x_1 )^{-1} ( 1 - x_2 )^{-1} ( 1 - x_1 x_2 )^{-1} \\
lim_{\lambda \rightarrow - \infty} f (\lambda ) &=&
( 1 + x_1 )^{-1} ( 1 + x_2 )^{-1} ( 1 + x_1 x_2 )^{-1}
\end{array}
\en
\end{lemma}

But now we have derived the signature characters for almost all
$A_2$ highest weight representations.

\begin{theorem}
\label{a2thm}
In the above notation, the normalised $A_2$ signature characters
for $V( \lambda )$ are given by
\eq
\sigma ( \lambda ) = f ( \lambda ) \, .
\en
\end{theorem}
\begin{proof}
This follows from theorem \ref{guessright} and lemmas \ref{l.asymp},
\ref{l.a2pmscde} and \ref{l.a2asy}.
\end{proof}

Thus we have signature characters for
$V( \lambda )$ for all $\lambda \in R$.
We proceed to deduce the signature characters for the remaining $A_2$
highest weight representations.
\begin{theorem}
If $\lambda$ lies on a vanishing line section bounding (only) the
two regions $R (a_1^{(1)} , a_2^{(1)} , a_3^{(1)} )$ and
$R (a_{1}^{(2)}, a_{2}^{(2)}, a_{3}^{(2)})$,
the normalised $A_2$ signature character for $V( \lambda )$ is
\eq
\sigma ( \lambda ) = \frac{
 \sum_{i=1}^2 \tilde{f} (a_1^{(i)} , a_2^{(i)} , a_3^{(i)} ) }
{ 2 (1-x_1^2 )(1 - x_2^2 )( 1 - (x_1 x_2 )^2 )} \, .
\en
\end{theorem}
\begin{proof}
This follows from lemma \ref{l.average} and theorem \ref{a2thm}.
\end{proof}

Now we calculate $\sigma(\lambda )$ for $\lambda$ lying
on the intersection of two or more vanishing lines.
Any such $\lambda$ clearly lies either on one of the $P(n, \alpha_1 )$ or on
one of the $P(n, \alpha_2 )$ (possibly both).
\begin{theorem}
\label{intersect}
If $\lambda$ lies on the intersection of two or more vanishing lines,
one of which is $P(n_i , \alpha_i )$ for $i=1$ or $2$ and some $n_i$,
let the regions that have $\lambda$ as a corner point and
a section of $P( n_i , \alpha_i )$ as an edge be
$R (a_1^{(j)} , a_2^{(j)} , a_3^{(j)})$ for $j = 1,2,3,4$.
Then
the normalised $A_2$ signature character for $V( \lambda )$ is
\eq
\label{e.pt}
\sigma ( \lambda ) = \frac{ \sum_{j=1}^{4}
 \tilde{f} (a_1^{(j)} , a_2^{(j)} , a_3^{(j)} ) }{
4 (1-x_1^2 )(1 - x_2^2 )( 1 - (x_1 x_2 )^2 )} \, .
\en
\end{theorem}

\begin{proof}
First let us
suppose that $\lambda_0$ lies on the intersection of two or more vanishing
lines, one of which is $P(n_1 , \alpha_1 )$ for some $n_1$.
If $\lambda_0$ lies on the intersection of precisely two vanishing lines,
denote the second vanishing line by $P( n_0 , \alpha_0 )$; if $\lambda_0$ lies
at a triple intersection, denote by $P( n_0 , \alpha_0 )$ the line
of the form $P(n_2 , \alpha_2 )$ on which $\lambda_0$ lies.

The following facts are known about the representation $V( \lambda_0 )$.
Besides the vector $\ket{ \lambda_0}$, $V ( \lambda_0 )$ contains highest
weight vectors
$v_0 \in V ( \lambda_0 )_{ n_0 \alpha_0 }$ and
$v_1 \in V ( \lambda_0 )_{ n_1 \alpha_1 } $.
Other than $v_0$, $V( \lambda_0 )$ contains no highest weight vectors
not lying in $U ( n_- ) v_1 $.  $U( n_- ) v_0 $ forms a representation of
$g$ isomorphic to $V( \lambda_0 - n_0 \alpha_0 )$.
We set
\eq
\tilde{P}( \mu ; \lambda_0 ) =
\dim \left( \frac{V ( \lambda_0 )_{\mu} \cap (U( n_- ) v_0 )}
{ V ( \lambda_0 )_{\mu} \cap (U( n_- ) v_0 ) \cap (U( n_- ) v_1 )} \right) \, .
\en

Define the fundamental weights $\Lambda_1 $ and $ \Lambda_2 $ by
$\iptwo{ \Lambda_i }{\alpha_j } = \delta_{ij} $; define coordinates
$(t, \eps )$
for $\lambda \in h^*$
by $\lambda = \lambda_0 + t \Lambda_1 + \eps \Lambda_2 $.
We write $V ( t , \epsilon )$ for $V( \lambda )$ and
$\ket{ t , \epsilon }$ for $\ket{\lambda }$, and we adopt the
convention that any vector written in the form $w ( t , \eps )$ belongs
to $V ( t , \epsilon )$.
Define the neighbourhood $N ( \delta )$ of
$\lambda_0$ as the coordinate region
with $ | t | < \delta$ and $  | \epsilon | < \delta$, and choose
$\delta_0 $ such that for non-zero $\eps < \delta_0 $ the representation
$V ( 0 , \eps )$ contains a unique descendent highest
weight vector $ v_1 ( 0 , \eps ) \in V (0 , \eps )_{n_1 \alpha_1}$;
set $ v_1 ( 0 , 0 ) = v_1$.
We can express $v_0 ( 0 , 0 )$ as $a \ket{ 0 , 0 }$, where
$a \in U( n_- )$.
Define the vectors $v_0 (0 , \eps ) \in V (0 , \eps )_{n_0 \alpha_0}$
by $v_0 ( 0 , \eps ) = a \ket{  0 , \eps}$.

If $\mu + n_1 \alpha_1 \notin \Lambda_- $ and
$\mu + n_0 \alpha_0 \notin \Lambda_- $ then
the inner product matrix $M_{\mu} (\lambda )$ is non-singular
in $N ( \delta_0 )$ and its signature is constant throughout the
neighbourhood.
Now we consider $M_{\mu} (\lambda )$ for $\mu$ such that
$\mu + n_1 \alpha_1 \in \Lambda_- $ or
$\mu + n_0 \alpha_0 \in \Lambda_- $.
Reducing $\delta_0$ if necessary, we choose
\eq
\begin{array}{ll}
a_i \in (U( n_- ))_{\mu + n_1 \alpha_1} & \for
1 \leq i \leq P( \mu + n_1 \alpha_1  ) \, ,  \\
b_i \in (U( n_- ))_{\mu + n_0 \alpha_0 }  & \for
    P( \mu + n_1 \alpha_1  ) + 1 \leq i \leq
     \tilde{P}( \mu ; \lambda_0 ) + P( \mu + n_1 \alpha_1  )  \, , \\
c_i \in (U( n_- ))_{\mu} & \for
\tilde{P}( \mu + \lambda_0 ) + P( \mu + n_1 \alpha_1  ) + 1 \leq i
                                  \leq P( \mu ) \, ,
\end{array}
\en
such
that, setting
\eq
w_i (0 , \epsilon ) = \left\{
\begin{array}{ll}
a_i v_1 ( 0 , \eps ) & \for  1 \leq i \leq P( \mu + n_1 \alpha_1  ) \, ,  \\
b_i v_0 ( 0 , \eps ) & \for P( \mu + n_1 \alpha_1  ) + 1 \leq i \leq
     \tilde{P}( \mu ; \lambda_0 ) + P( \mu + n_1 \alpha_1  )  \, , \\
c_i \ket{ 0 , \eps }  & \for
\tilde{P}( \mu + \lambda_0 ) + P( \mu + n_1 \alpha_1  ) + 1 \leq i
                                  \leq P( \mu ) \, ,
\end{array}
\right.
\en
the set $\{ w_i ( 0 , \epsilon ) : 1 \leq i \leq P ( \mu ) \}$ forms
a basis of $V ( 0 , \eps )_{\mu}$ for $\eps \leq \delta_0 $.

In this basis the inner product matrix $M_{\mu} ( 0 , \eps )$ has the form
\eq
 \left( \begin{array}{ccc}
           0  &          0                       &   0 \\
           0  &          \eps M' ( \eps )         &  \eps A ( \eps ) \\
           0  &      \eps A^T (\eps )            &  M'' (\eps )
          \end{array} \right)   \, .
\en
We have that $M'' (\eps ) = M'' ( 0 ) + O( \eps )$ and
we know that $M'' ( 0 )$ is non-singular, since if it were not
$V( 0 , 0 )$ would contain a highest weight vector not descended from
$v_0$ or $v_1$.
Hence there is a new basis in which $M_{\mu} (0,  \eps )$ has the form
\eq
 \left( \begin{array}{ccc}
           0  &          0                       &   0      \\
           0  & \eps M' ( 0) +   O( \eps^2 )     &   0      \\
           0  &     0            &  M'' (0 ) + O( \eps )
          \end{array} \right)   \, .
\en
Now a generalisation of the
Shapovalov formula obtained by one of us (A.K.)
implies that the submatrix
\eq
 \left( \begin{array}{cc}
            \eps M' ( 0) +   O( \eps^2 )     &   0      \\
              0            &  M'' (0 ) + O( \eps )
          \end{array} \right)   \, .
\en
has determinant of order $ \eps^{\tilde{P}( \mu ; \lambda_0 )}$.
(For completeness, a proof of this last result is given in an appendix.)
Hence $M' ( 0 )$ must be non-singular, and we have that
\eq
\sigma ( 0 , 0 ) = \frac{1}{2}
            ( \lim_{\eps \rightarrow 0^+ } \sigma ( 0 , \eps ) +
              \lim_{\eps \rightarrow 0^- } \sigma ( 0 , \eps ) )
\en
Equation (\ref{e.pt}) follows.

When $\lambda$ lies on the intersection of two or more vanishing
lines, one of which is $P(n_2 , \alpha_2 )$ for some $n_2$, the argument
is similar.  This completes the proof.
\end{proof}

The case when $\lambda$ lies on a triple intersection of vanishing lines
is particularly interesting.
Here, theorem \ref{intersect} gives us two
expressions for $\sigma ( \lambda )$.
Moreover, since these representations are unitary, the (unnormalised)
signature character equals the ordinary character, which
has a well-known expression given by the Weyl character formula.
However, it is easy to see that the three expressions are equal.
Explicitly, if $\lambda = (r_1 \Lambda_1 + r_2 \Lambda_2 )$, for
non-negative integers $r_1$ and $r_2$, then theorem \ref{intersect}
implies that
\eq
\begin{array}{rcl}
\sigma( \lambda ) &=& (1 - x_1 )^{-1} (1 - x_2 )^{-1}(1 -x_1 x_2)^{-1}\\
&& ~\times
 ( 1 - x_1^{r_1 + 1} - x_2^{r_2 + 2} + x_1^{r_1 +1} (x_1 x_2 )^{r_2 + 1}\\
&&~+ x_2^{r_2 +1} (x_1 x_2 )^{r_1 + 1} - (x_1 x_2 )^{r_1 + r_2 + 2} )
\end{array}
\en
This gives us an alternative proof of the well known result that
$V(\lambda )$ is unitary if $\lambda = (r_1 \Lambda_1 + r_2 \Lambda_2 )$, for
non-negative integers $r_1$ and $r_2$.

\begin{corollary}
Let $\lambda = (r_1 \Lambda_1 + r_2 \Lambda_2 )$ for
non-negative integers $r_1$ and $r_2$, and
let $V' ( \lambda )$ be the maximal proper submodule of $V( \lambda )$.
Then the bilinear form $\ipone{~}{~}'$
induced on the representation $V (\lambda) / V' ( \lambda )$ by
$\ipone{~}{~}$ is positive definite.
(In other words, the irreducible representation with
highest weight $\lambda$ is unitary.)
\end{corollary}
\begin{proof}
The quotient module $V' ( \lambda )$ is the submodule of $V( \lambda )$
generated by all the states $v \in V( \lambda )$ such that
$\ipone{v}{w} = 0$ for all $w \in V( \lambda )$.
So the representation $V (\lambda) / V' ( \lambda )$ has the
same signature character with respect to $\ipone{~}{~}'$
as $V( \lambda )$ does with respect to  $\ipone{~}{~}$.
In other words,
\eq
\begin{array}{rcl}
\chi^{\rm sig} ( V (\lambda) / V' ( \lambda ) ) &=&
(1 - x_1 )^{-1} (1 - x_2 )^{-1} (1 - x_1 x_2 )^{-1} \\
&& ~\times x_1^{r_1} x_2^{r_2}
 (1-x_1^{r_1 +1}- x_2^{r_2 + 2} + x_1^{r_1 +1} (x_1 x_2 )^{r_2 + 1}\\
&& \qquad + x_2^{r_2 +1} (x_1 x_2 )^{r_1 + 1} - (x_1 x_2 )^{r_1 + r_2 + 2} ) \\
&=&  \chi ( V (\lambda) / V' ( \lambda ) ) \, ,
\end{array}
\en
the last equality following from the Weyl character formula.  This completes
the proof.
\end{proof}

Finally, we note that the signature characters on $R$
can be re-expressed in a suggestively simple way.
Denote the Weyl reflection corresponding to $\alpha \in \Delta_{+}$ by
$r_{\alpha}$ and let $C_0 \equiv C$ be the fundamental Weyl chamber
for $A_2$.
Denote the other Weyl chambers for $A_2$ by
$C_1 = r_{\alpha_1} C$, $C_2 = r_{\alpha_2} C$,
$C_3 = r_{\alpha_1} r_{\alpha_2} C $,
$C_4 = r_{\alpha_2} r_{\alpha_1} C$,
$C_5 = r_{\alpha_3} C$.
Then
\eq
f ( \lambda ) =
\frac{\tilde{f_i} (a_1 , a_2 , a_3 )}
{(1-x_1^2 )(1 - x_2^2 )( 1 - (x_1 x_2 )^2 )}
\en
for $\lambda \in R ( a_1 , a_2 , a_3 )$ such that
$\lambda + \rho \in C_i$,
where
\eq
\begin{array}{rcl}
\tilde{f}_0  ( a_1, a_2, a_3 ) &=&  (1+x_1 ) (1 + x_2 ) (1 + x_1 x_2 )\\
&&-          2 x_1^{a_1 +1} ( 1 + x_2 ) (1 + x_1 x_2)         \\
&&-             2 x_2^{a_2 + 1} ( 1 + x_1 ) (1 + x_1 x_2)     \\
&&+  4 x_1^{a_1 + 1} (x_1 x_2 )^{a_2 + 1} ( 1 + x_2 ) \\
&&+ 4 x_2^{a_2 + 1} (x_1 x_2 )^{a_1 + 1} ( 1 + x_1 )  \\
&&- 8 ( x_1 x_2 )^{a_1 + a_2 + 2 } \\
&&+ 2 (x_1 x_2 )^{a_3 + 1} ( 1 - x_1 ) ( 1 - x_2 )        \\
\tilde{f}_1 ( a_1 , a_2 , a_3 )&=& (1-x_1 )( 1 + x_2 )( 1 + x_1 x_2 )\\
&& -                   2 x_2^{a_2 + 1} ( 1 - x_1 x_2 ) ( 1 + x_1 )   \\
&& -                   2 (x_1 x_2)^{a_3 + 1} ( 1 - x_1 ) ( 1 + x_2 )  \\
&& +                   4 x_1^{a_3 + 1} x_2^{a_2 + 1} ( 1 - x_1 x_2 ) \\
\tilde{f}_2   ( a_1 , a_2 , a_3 ) &=&(1 - x_2 )(1 + x_1 )(1 + x_1 x_2)\\
&& -                2 x_1^{a_1 + 1} ( 1 - x_1 x_2 ) ( 1 + x_2 )   \\
&& -                   2 (x_1 x_2 )^{a_3 + 1} ( 1 - x_2 ) ( 1 + x_1 )  \\
&& +                   4 x_2^{a_3 + 1} x_1^{a_1 + 1} ( 1 - x_1 x_2 ) \\
\tilde{f}_3 ( a_1 , a_2 , a_3 ) &=& ( 1 - x_1 ) (1 - x_1 x_2 ) (1+x_2)\\
&& -                    2 x_2^{a_2 + 1} ( 1 - x_1 ) ( 1 - x_1 x_2 )   \\
\tilde{f}_4 ( a_1 , a_2 , a_3 ) &=&  ( 1 - x_2 ) (1 - x_1 x_2 ) ( 1 + x_1 )\\
          &&   -      2 x_1^{a_1 + 1} ( 1 - x_2 ) ( 1 - x_1 x_2 )   \\
\tilde{f}_5
( a_1 , a_2 , a_3 ) &=& (1 - x_1 )(1- x_2)( 1 - x_1 x_2)
\, .
\end{array}
\en
\section{Signature characters of $B_2$}

The $B_2$ Cartan matrix and symmetrised Cartan matrix are given by
\eq
A = \left( \begin{array}{rr} 2 & -2 \\ -1 & 2 \end{array}\right) \,,\;
A^{sym} =\left( \begin{array}{rr}  1 & -1 \\ -1 & 2 \end{array}\right) \,.
\en
With this choice the simple roots are $\alpha_1,\alpha_2$ with
$\alpha_1$ the short root. The other positive roots are
$\alpha_3 = \alpha_1+\alpha_2$ and $\alpha_4 = 2\alpha_1+\alpha_2$.
The Weyl group of $B_2$ is the dihedral group $D_4$ and there are thus
eight Weyl chambers. We denote these as follows.
\eq
\begin{array}{rclrcl}
C_0 &=& C \, , \qquad\qquad\qquad & C_1 &=& r_1 C \, , \\
C_2 &=& r_1r_2C       \, , & C_3 &=& r_1r_2r_1C \, , \\
C_4 &=& r_2r_1r_2r_1C \, , & C_5 &=& r_2r_1r_2C \, , \\
C_6 &=& r_2r_1C       \, , & C_7 &=& r_2 C \, .
\end{array}
\en
The element $\rho\in h^*$ is given by
$\rho = \frac{1}{2}(4 \alpha_1+3\alpha_2 )$.
The region R splits into a union of connected subspaces
\eq
R = \cup R(a_1,a_2,a_3,a_4)
\en
where
\eq
R(a_1,a_2,a_3,a_4) = \{ \lambda\in h^* : a_i =
\max(0,\floor{ \frac{2(\alpha_i,\lambda+\rho)}{(\alpha_i , \alpha_i )}}) \}
\en
If $\lambda\in R(a_1,a_2,a_3,a_4)$ and $\lambda + \rho \in C_i$ then the
normalised signature character of $V(\lambda)$ is given by
\eq
\sigma ( \lambda ) = \frac{\tilde f_i(a_1,a_2,a_3,a_4)}
{(1 - x_1^2)(1 - x_2^2)(1 - (x_1x_2)^2)(1 - (x_1^2x_2)^2)} \, ,
\en
where the $\tilde{f}_i$ are defined as follows.

\eq
\begin{array}{rcl}
\tilde{f}_0 (a_1 , a_2 , a_3 , a_4 ) &=&
          ( 1 + x_1)(1+x_2)(1+x_1x_2)(1+x_1^2x_2) \\
	&& -2 x_1^{a_1+1}(1+x_2)(1+x_1x_2)(1+x_1^2x_2) \\
	&& - 2 x_2^{a_2+1}(1+x_1)(1+x_1x_2)(1 + x_1^2x_2) \\
	&& +2(x_1x_2)^{a_3+1}(1+x_1)(1-x_2)(1-x_1^2x_2) \\
	&& +2(x_1^2x_2)^{a_4+1}(1-x_1)(1+x_2)(1-x_1x_2) \\
	&& -4 (x_1^2x_2)^{a_4+1}(x_2)^{a_2 + 1}(1-x_1)(1-x_1x_2) \\
	&& -4 (x_1x_2)^{a_3+1}x_1^{a_1+1}(1-x_2)(1-x_1^2x_2) \\
	&& +4 (x_1x_2)^{a_1+1}x_2^{a_2+1}(1+x_1)(1+x_1^2x_2) \\
	&& +4 x_1^{a_1+1}(x_1^2x_2)^{a_2+1}(1+x_2)(1+x_1x_2) \\
	&& -8 (x_1x_2)^{2a_2 + a_1 + 3}(1 + x_1) \\
	&& -8 (x_1^2x_2)^{a_1+a_2+2}(1+x_2) \\
	&& +16 (x_1x_2)^{2 a_2+2}(x_1^2x_2)^{a_1+1} \, ,
\end{array}
\en
\eq
\begin{array}{rcl}
\tilde{f}_1 (a_1 , a_2 , a_3 , a_4 ) &=&
         (1-x_1)(1+x_2)(1+x_1x_2)(1+x_1^2x_2) \\
	&& -2 x_2^{a_2+1}(1+x_1)(1-x_1x_2)(1+x_1^2x_2) \\
	&& -2 (x_1^2x_2)^{a_4+1}(1-x_1)(1+x_2)(1+x_1x_2) \\
	&& +2 (x_1x_2)^{a_3+1}(1-x_1)(1-x_2)(1-x_1^2x_2) \\
	&& +4 x_2^{a_2+1}(x_1^2x_2)^{a_4+1}(1+x_1)(1-x_1x_2) \\
	&& +4 (x_1x_2)^{a_2+1}(x_1)^{a_4+1}(1+x_2)(1-x_1^2x_2) \\
	&& -8(x_1x_2)^{a_2+a_4+2}(1 - x_1^2x_2) \, ,
\end{array}
\en
\eq
\begin{array}{rcl}
\tilde{f}_2 (a_1 , a_2 , a_3 , a_4 ) &=&
          (1-x_1)(1+x_2)(1+x_1x_2)(1-x_1^2x_2) \\
	&& -2 x_2^{a_2+1}(1+x_1)(1-x_1x_2)(1-x_1^2x_2) \\
	&& -2 (x_1x_2)^{a_3+1}(1-x_1)(1+x_2)(1-x_1^2x_2) \\
	&& +4 x_1^{a_3+1} x_2^{a_2+1}(1 - x_1x_2)(1-x_1^2x_2) \, ,
\end{array}
\en
\eq
\begin{array}{rcl}
\tilde{f}_3 (a_1 , a_2 , a_3 , a_4 ) &= &
          ( 1+x_2 )(1-x_1)(1-x_1x_2)(1-x_1^2x_2) \\
          && - 2 x_2^{a_2+1}(1-x_1)(1-x_1x_2)(1-x_1^2x_2)  \, ,
\end{array}
\en
\eq
\tilde{f}_4 (a_1 , a_2 , a_3 , a_4 ) =
          (1-x_1)(1-x_2)(1-x_1x_2)(1-x_1^2x_2) \, ,
\en
\eq
\begin{array}{rcl}
\tilde{f}_5 (a_1 , a_2 , a_3 , a_4 ) & = &
          (1 + x_1 )(1-x_2)(1-x_1x_2)(1-x_1^2x_2) \\
           &&  - 2 x_1^{a_1+1}(1-x_2)(1-x_1x_2)(1-x_1^2x_2) \, ,
\end{array}
\en
\eq
\begin{array}{rcl}
\tilde{f}_6 (a_1 , a_2 , a_3 , a_4 ) &=&
           (1+x_1)(1-x_2)(1-x_1x_2)(1+x_1^2x_2) \\
	&& -2 x_1^{a_1+1}(1+x_2)(1-x_1x_2)(1-x_1^2x_2) \\
	&& -2 (x_1^2x_2)^{a_4+1}(1+x_1)(1-x_2)(1-x_1x_2) \\
	&& +4 x_1^{a_1+1}x_2^{a_4+1}(1-x_1x_2)(1-x_1^2x_2) \, ,
\end{array}
\en
and
\eq
\begin{array}{rcl}
\tilde{f}_7 (a_1 , a_2 , a_3 , a_4 ) &=&
           ( 1 + x_1 ) ( 1 - x_2 ) ( 1 + x_1 x_2 ) ( 1 + x_1^2 x_2 ) \\
        && -2 x_1^{a_1 + 1} ( 1 + x_2 ) ( 1 + x_1 x_2 ) ( 1 - x_1^2 x_2 ) \\
        && +2 (x_1^2 x_2 )^{a_4 + 1} ( 1 - x_2 ) ( 1 - x_1 ) ( 1 - x_1 x_2 ) \\
        && -2 (x_1 x_2 )^{a_3 + 1} ( 1 + x_1 ) (1 - x_2 ) ( 1 + x_1^2 x_2) \\
        && +4 x_1^{a_1+1} (x_1 x_2 )^{a_3 + 1} ( 1 + x_2) ( 1 - x_1^2 x_2 ) \\
        && + 4 x_2^{a_5 + 1} x_1^{a_1 + 1} (1+ x_1 ) ( 1 - x_1 x_2 )
               (1 + x_1 x_2^{\eps} ) \\
        && - 8 x_2^{a_5 + 1} x_1^{a_1 + a_3 + 2} ( 1 - x_1 x_2 )
                ( 1 + x_1 x_2^{\eps}) \, ,
\end{array}
\en
where
\eq
a_5 = \floor{\frac{a_1 + a_3 + 1}{2}}
\en
and
\eq
\eps = 2 a_5 - a_1 - a_3 - 1 \, .
\en

\section{Acknowledgements}
A.K. thanks SERC for an Advanced Fellowship and
Sidney Sussex College, Cambridge for the
Knox-Shaw Research Fellowship.
G.W. thanks SERC for a Research Studentship and a
Research Assistantship under grant GR/F/72192,
and Trinity College, Cambridge
for a Rouse-Ball Travelling Studentship.
This work was supported in part by DOE grant DE-AC02-76ER02220 at the
Princeton
Institute for Advanced Study and by funds from the US National Science
Foundation under grant number PHY89-04035.
We are grateful for the hospitality of the
Santa Barbara Institute for Theoretical Physics.
We thank D. Friedan, \mbox{P. Goddard,} Z. Qiu and S. Shenker for many helpful
discussions.

\newpage
\appendix
\section{Appendix}
Here we establish formulae for the determinants of certain submatrices of the
inner product matrices of $A_2$ highest weight representations.
We consider the inner product matrix $M_{\mu} (\lambda )$ on
$V( \lambda )_{\mu}$ , where
$\mu = - (m_1 \alpha_1 + m_2 \alpha_2 )$ is an $A_2$
weight, in the basis
$\{ v_i (\lambda ) : 1 \leq i \leq \min( m_1 ,m_2 ) + 1 \}$,
where
\eq
v_i ( \lambda ) = ( E_{- \alpha_1 } )^{m_1 - i+1}
                  ( E_{-\alpha_1 - \alpha_2})^{i-1}
                  ( E_{- \alpha_2 } )^{m_2 -i+1} \ket{\lambda} \, .
\en
For $r_1 $ and $r_2 $ positive integers with $r_1  \leq m_1 $ and
$r_2  \leq m_2$, define the
submatrices
$M_{\mu}^{\alpha_1 , r_1 } (\lambda )$ and
$M_{\mu}^{\alpha_2 , r_2 } (\lambda )$ of $M_{\mu} (\lambda )$
by
\eq
\begin{array}{rcl}
( M_{\mu}^{\alpha_1 , r_1 } (\lambda ) )_{ij} &=&
\ipone{ v_{m_1 -r_1 +1+i}}{ v_{m_1 -r_1 +1+j} }  \\
&& \for 1 \leq i,j \leq \min ( m_1 ,m_2 ) - m_1 + r_1 \, ,\\
( M_{\mu}^{\alpha_2 , r_2 } (\lambda ) )_{ij} &=&
\ipone{ v_{m_2 -r_2 +1+i}}{ v_{m_2 -r_2 +1+j} }  \\
&& \for 1 \leq i,j \leq \min ( m_1 ,m_2 ) - m_2 + r_2 \, ,
\end{array}
\en
with the convention that the matrices are
null if the ranges of $i$ and $j$ are empty.

The result quoted in the proof of theorem \ref{intersect} is a
corollary of the following.

\begin{lemma}
In the above notation, consider
$ M_{\mu}^{\alpha_1 , r_1 } (r_1 \Lambda_1 + r_2 \Lambda_2 )$, for
fixed positive integer $r_1 $, as a matrix of polynomials in $r_2 $.
Similarly, consider
$ M_{\mu}^{\alpha_2 , r_2 } (r_1 \Lambda_1 + r_2 \Lambda_2 )$, for
fixed positive integer $r_2 $, as a matrix of polynomials in $r_1 $.  Then
\eq
\label{subdet1}
\begin{array}{rcl}
\det ( M_{\mu}^{\alpha_1 , r_1 } ((r_1 - 1) \Lambda_1 + r_2 \Lambda_2  ) ) &=&
C_1 ( \prod_{p=1}^{m_2 } ( r_2  - p + 1)^{A(m_1 ,m_2 ,r_1 ,p)} ) \\
&&~~(\prod_{p=1}^{r_1 -1} ( r_2  + p )^{B(m_1 ,m_2 ,r_1 ,p)} )
\end{array}
\en
and
\eq
\label{subdet2}
\begin{array}{rcl}
\det ( M_{\mu}^{\alpha_2 , r_2 } (r_1 \Lambda_1 + (r_2 - 1) \Lambda_2  ) ) &=&
C_2 ( \prod_{p=1}^{m_1} ( r_1  - p + 1 )^{A (m_2 ,m_1 ,r_2 ,p)} ) \\
&& ~~ ( \prod_{p=1}^{r_2 - 1} ( r_1  + p )^{B (m_2 ,m_1 ,r_2 ,p)} )
\end{array}
\en
where $C_1$ and $C_2$ are non-zero constants and
\eq
\begin{array}{rcl}
A(m_1 ,m_2 ,r_1 ,p) &=&
\max ( \min( m_1, m_2 -p) + 1,0 ) \\
&& - \max ( \min ( m_1 - r_1 - p , m_2 - p) + 1,0) \\
&& - \max ( \min ( m_1 -r_1 , m_2 - r_1 - p) +1,0) \\ && +
\max ( \min ( m_1 - r_1 - p, m_2 - r_1 - p)+1,0) \, , \\ && \\
B(m_1 ,m_2 ,r_1 ,p) &=&
\min ( \max ( m_1 - p + 1,0), \max ( m_2 - p + 1, 0) ) \\
&& - \min( \max ( m_1 - r_1 + 1,0), \max (m_2 - p + 1,0)) \, .
\end{array}
\en
\end{lemma}
\begin{proof}
We prove equation (\ref{subdet1}); the proof of (\ref{subdet2})
is similar.
Fix $r_1 $ and
consider $\lambda$ of the form $(r_1 - 1) \Lambda_1 + r_2 \Lambda_2 $
(for any $r_2 $).
The vector $ v(\lambda) = ( E_{-\alpha_1} )^{r_1} \ket{\lambda}$ is a
highest weight vector in $V( \lambda )$; it generates a submodule
$U(n_- ) v(\lambda )$ of $V( \lambda )$; we write
$U_{\mu} (\lambda ) = U(n_- ) v(\lambda ) \cap (V (\lambda ))_{\mu}$.
Now the submodule embeddings of the modules
$V( \lambda )$ are known to be such that
for $r_2  = p-1$ and $p$ a positive integer with
$ 1 \leq p \leq m_2$
the null space of $M_{\mu} (\lambda )$ has dimension
$\dim ( U_{\mu} (\lambda )) + A(m_1 ,m_2 ,r_1 ,p )$; that for
$r_2  = -p$ and $p$ a positive integer with $1 \leq p \leq (r_1 - 1)$ the
null space of $M_{\mu} (\lambda )$ has dimension
$\dim ( U_{\mu} (\lambda )) + B(m_1 ,m_2 ,r_1 ,p )$; and that
for all other $r_2 $
the null space of $M_{\mu} (\lambda )$ has dimension
$\dim ( U_{\mu} (\lambda ))$.
The vectors in the set
$\{ v_i (\lambda ) :  m_1 - r_1 + 2 \leq i \leq \min(m_1 ,m_2 ) + 1 \}$
span
a subspace of $(V (\lambda ))_{\mu}$ complementary to $U_{\mu} (\lambda )$,
and $M_{\mu}^{\alpha_1 , r_1 } ( \lambda )$ is the matrix of their inner
products.
Let $p$ be a positive integer with $1 \leq p \leq m_2$.
Then we have that
$M_{\mu}^{\alpha_1 , r_1 } ((r_1 - 1) \Lambda_1 + r_2 \Lambda_2   )$
has a null space of dimension
$A(m_1 ,m_2 ,r_1 ,p )$ at $r_2  = (p-1)$.
So $\det ( M_{\mu}^{\alpha_1 , r_1 } ((r_1 - 1) \Lambda_1 +
(p-1+\eps ) \Lambda_2 ))$
is at least of order
$\eps^{A(m_1 ,m_2 ,r_1 ,p)}$.
Thus $\det ( M_{\mu}^{\alpha_1 , r_1 } ((r_1 - 1) \Lambda_1 +
r_2 \Lambda_2   ) )$
is divisible by $( r_2  - p+ 1 )^{A(m_1 ,m_2 ,r_1 ,p)}$.
A similar argument shows that it is also divisible by
$( r_2  + p )^{B(m_1 ,m_2 ,r_1 ,p)}$ if $p$ is a positive integer with
$1 \leq p \leq (r_1 - 1)$.
Hence the polynomial
\eq
( \prod_{p=1}^{m_2 } ( r_2  - p+1)^{A(m_1 ,m_2 ,r_1 ,p)} )
  ( \prod_{p=1}^{r_1 -1} ( r_2  + p )^{B(m_1 ,m_2 ,r_1 ,p)} )
\en
divides
\mbox{$\det ( M_{\mu}^{\alpha_1 , r_1 } ((r_1 - 1) \Lambda_1 +
(p-1)\Lambda_2 ))$.}
Now the order of the determinant as a polynomial in $r_2 $ is at most
$\max ( m_2 ( \min ( m_1,m_2 ) - m_1+ r_1 ), 0 )$, since this is
the order of the
product of the diagonal elements, which dominates all other contributions.
But it is easy to verify that
\begin{eqnarray}
\lefteqn{
( \sum_{p=1}^{m_2 } A(m_1 ,m_2 ,r_1 ,p) ) +
  ( \sum_{p=1}^{r_1 -1} B(m_1 ,m_2 ,r_1 ,p) ) = ~~~~~~~~~~~~~~ } \\
& & ~~~~~~~~~~~\max ( m_2 ( \min ( m_1 ,m_2 ) - m_1 + r_1 ), 0 ) \nonumber \, .
\end{eqnarray}
Thus equation (\ref{subdet1}) must hold for some constant $C_1$.
Finally, since the matrix
\mbox{$ M_{\mu}^{\alpha_1 , r_1 } ((r_1 -1) \Lambda_1 + r_2  \Lambda_2 )$} is
non-singular for non-integer $r_2 $, $C_1$ must be
non-zero.
\end{proof}

\end{document}